\documentclass[prb,manuscript,showpacs,superscriptaddress,floatfix]{revtex4}
\usepackage{amssymb,mathptm,multirow,makecell,array,ragged2e}
\usepackage{amsmath,amssymb,amsfonts,bm}
\usepackage{graphicx,epsfig,color}
\usepackage{latexsym}
\usepackage{ulem}
\usepackage{booktabs}

\begin{document}

\title{Nanopatterning graphite by ion-beam-sputtering: Effects of polycrystallinity}

\author{Sun Mi Yoon}
\affiliation{Department of Physics, Sook-Myung Women's University, Seoul 140-742, Korea}
\author{J.-S. Kim}
\email[]{jskim@sm.ac.kr}
\affiliation{Department of Physics, Sook-Myung Women's University, Seoul 140-742, Korea}
\author{D. Yoon}
\affiliation{Department of Physics, Sogang University, Seoul 121-742, Korea}
\author{H. Cheong}
\affiliation{Department of Physics, Sogang University, Seoul 121-742, Korea}
\author{Y. Kim}
\affiliation{Beamline Research Division, Pohang Accelerator Laboratory (PAL), Pohang 790-784, Korea}
\author{H. H. Lee}
\affiliation{Beamline Research Division, Pohang Accelerator Laboratory (PAL), Pohang 790-784, Korea}

\date{Received \today }

\begin{abstract}

 Employing graphites having distinctly different mean grain sizes, 
 we study the effects of polycrystallinity on the pattern formation by ion-beam-sputtering.  
 The grains influence the growth of the ripples in highly anisotropic fashion; 
 Both the mean uninterrupted ripple length along its ridge and the surface width depend on the mean size of the grains, 
 which 
is attributed to the large sputter yield at the grain boundary compared with that on terrace.
 In contrast, the ripple wavelength does not depend on the mean size of the grains. 
Coarsening of the ripples-accompanying the mass transport across the grain boundaries-should not be driven 
by thermal diffusion, rather by ion-induced processes.    
\end{abstract}


\maketitle

\section{Introduction}

Incidence of energetic ion-beam causes the displacement and even removal of atoms near the surface. 
These whole processes are conventionally referred to as ion-beam-sputtering (IBS). 
Simultaneously, the healing kinetics proceeds via mass transport to minimize the surface free energy of the modified surface. Those two competing processes often produce patterns of nano-dots/holes or ripples depending on the incidence angle of the ion-beam. IBS is one of the most versatile tools to fabricate nano patterns in various sizes and shapes by controlling physical variables, and applicable to a wide range of materials from metals, insulators to organic materials. IBS has, thus, drawn attention as a representative method for triggering physical self-assembly.\cite {W.L.Chan2007,JPCD special issue}

Continuum models have mostly elucidated the pattern formation by IBS.  
In their seminal work, Bradley and Harper (BH)\cite{BH} proposed a model of the pattern evolution; They took the erosion into account according to Sigmund picture\cite{Sigmund} in the linear approximation, while the diffusion of adatoms according to Mullin's model.\cite{Mullin} According to Sigmund, the sputter yield at a position on the surface is proportional to the energy deposited at that position by an incident ion. The energy depends on Gaussian ellipsoidal function of the distance from a terminal position of an incident ion to the position at the surface.\cite{Sigmund} According to Mullin's picture, the adatom current is proportional to the gradient of the surface free energy which is in proportion to the curvature at the position of interest .\cite{Mullin} 
BH model and its non-linear extensions\cite{KS,eKS,eKS-Cuerno,dKS} that incorporate surface-confined mass flow, redeposition and surface damping well reproduce, although qualitatively yet, many features of the pattern formation by IBS.  
Those models, though, tacitly assume amorphous surface. 
Refined models have also been proposed taking the crystallinity\cite{crystallinity} of the substrate and anisotropy\cite{anisotropy} of the surface and sputter geometry  combined into account. 


 Polycrystals consist of the grains 
whose mean size can be comparable with characteristic size of the structures formed by IBS.\cite{D.Ghose2009,deMongeot}
Polycrystals offer unique environment for the pattern formation by IBS contrasted with the homogeneous substrates such as the amorphous and the single crystals. 
 A challenging question is, then, whether and in the case how the grained structure of the polycrystalline substrate affects the pattern formation and its temporal evolution. 
 
 Recently, Toma {\it et al.}\cite{deMongeot} have studied the ripple evolution of polycrystalline Au films by IBS.
With continued sputtering, the initially rough surface leveled off, and the pattern evolved as if it did on single crystalline Au surface in regards to the surface width and ripple wavelength. This observation leads them to the conclusion that the polycrystallinity did not influence the pattern evolution. On the other hand, \v{S}kere\v{n} {\it et al.}\cite{Skeren} reported that the variation of the sputter yield depending on the orientation of the grains induced the surface instability during IBS, and well reproduced the morphological evolution of polycrystalline Ni films without invoking the instability owing to the curvature dependent erosion\cite{BH}.  
The two conclusions on the effects of the polycrystallinity contradict to each other. 

Highly oriented pyrolytic graphite (HOPG) is polycrystalline
and composed of the large grains, compared with the metallic films such as Au\cite{deMongeot}  and Ni\cite{Skeren}.
HOPG would, thus, offer the opportunity to study the effects of the grained structure on the pattern formation in the limit opposite to that of the metallic films.  Several groups have previously worked on patterning HOPG by IBS.\cite{Takahiro2007,Takahiro2009,Habenicht1999,Habenicht2001}  
Their studies are, however, little motivated by the interests in the effects of polycrystallinity on pattern formation.

In this work, we study the effects of the grained structure on the pattern formation, employing two different kinds of graphites, HOPG and natural graphite (NG). The mean grain size of NG is distinctly larger than that of HOPG. 
This controlled experiment clearly tells that the grain boundaries play a critical role in determining the mean uninterrupted length of the ripples along their ridges or the coherence length $\ell$ and the surface width $W$, while they little influence the ripple wavelength.
 This highly anisotropic effects on the ripple evolution are attributed to the intricate roles of the grain boundaries in the temporal evolution of the primordial islands to the ripples during the pattern formation. 
   

\section{Experiments}

The ion-beam-sputtering of both HOPG (ZYA grade, SPI) and NG (donated from Union Carbide) samples were performed in a high vacuum chamber whose base pressure was 5$\times$ 10$^{-9}$ Torr. The ion-irradiated surface is characterized {\it ex situ} by  atomic force microscopes (AFM) in both the contact (AFM, PSI, Autoprobe CP) and the noncontact modes (XEI-100, Park Systems). 
Sputtering is performed by irradiation of Ar$^+$ ion-beam with its beam diameter $\sim10$ mm, incident ion energy $E_{ion}$, 2 keV, at a polar angle of incidence $\theta$, 78$^{\circ}$ from the global surface normal. The partial pressure of Ar$^+$, $P_{Ar}$ and the ion flux $f$ are 1.2 $\times$ 10$^{-4}$ Torr and 0.3 ions nm$^{-2}$ s$^{-1}$, respectively.
The sample temperature is kept around room temperature by limiting each sputter period to 1 minute with interval for 10 minutes, unless mentioned otherwise. 


The Raman spectra of the samples were obtained by using a micro-Raman spectroscopy system.\cite{Yoon2009,YoonPRB2009} The 514.5 nm (2.41 eV) line of an Ar ion laser was used as the excitation source and the laser power was kept below 1mW to avoid unintentional heating. The laser beam was focused-spot size $< 1 \mu$m-onto the graphite sample by a x 50 microscope objective lens (0.8 N.A.), and the scattered light was collected and collimated by the same objective. The collected Raman scattered light was dispersed by a Jobin-Yvon Triax 550 spectrometer and detected by a liquid-nitrogen-cooled charge-coupled-device detector.  The spectral resolution was about 1 cm$^{-1}$. 

The grazing incidence x-ray diffraction (GID) of the samples was performed with 20 keV photons ($\lambda$ = 0.62 $\AA$) in the 5A beam line of Pohang light source in Korea. The incident angle of the x-ray was kept to be ~ 0.1$^{\circ}$ from the global sample plane to reduce both the beam penetration depth and bulk diffuse scattering. 

\section{Results}

 \begin{figure*}[!t]
\begin{center}
\includegraphics [angle=0,width=0.8\textwidth]{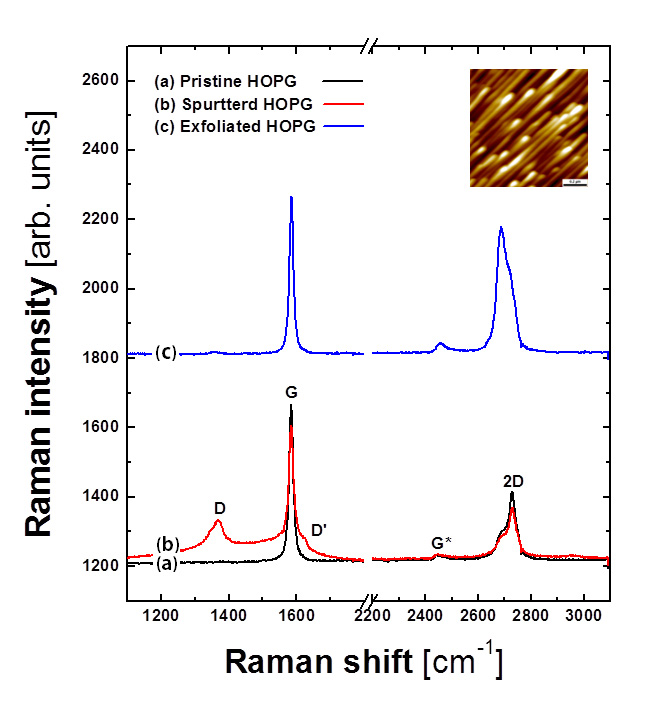}
\caption{Raman spectra (a) before and (b) after sputtering  HOPG with $\psi = 5625$ ions nm$^{-2}$. (c) Raman spectrum of the surface layers  peeled off from the HOPG giving spectrum (b). Inset: A nano ripple pattern corresponding to the spectrum (b). }
\end{center}
\end{figure*}

 In Fig. 1(a), the Raman spectrum of HOPG before IBS shows three bands, labeled as G, G* and 2D, typical of $pristine$ HOPG.\cite{Dresselhaus} 
After an extended sputtering with an ion fluence $\psi$ , flux times total sputter time, $\psi$ = 5625 ions nm$^{-2}$,  
its surface is patterned by ripples as shown in the inset of Fig. 1. Two new bands D and D' develop as shown in Fig. 1(b), while both G and 2D bands notably weaken as previously reported.\cite{Takahiro2007, Takahiro2009} Both D and D' bands originate from defective carbon atoms.\cite{Ferrari, Dresselhaus} The observed spectral changes tell that sputtering produces defective carbons at the cost of the $pristine$ carbons.

The sample, however, largely remains crystalline; The spectral shape of the {\it amorphous} carbon shows broad adjoined peaks of  D and G bands,\cite{Robertson} while the present spectrum in Fig. 1 shows a well-defined D band with still intense and sharp G band.
 Since the Raman spectrum could also sample the $pristine$ layers beneath the damaged surface layers, we exfoliated the surface layers of the HOPG after  transferring it to silicon oxide substrate. Fig. 1(c) is a Raman spectrum of the exfoliated surface layers that should be thinner than 10 layers, because the intensity of the 2D band is now comparable with that of the G band.\cite{Yoon2009} In Fig. 1(c), we still find a sharp G band with negligible D band, confirming the crystallinity of the sputtered surface. 
 Takahiro {\it et al.}\cite{Takahiro2007, Takahiro2009} also observed the development of both D and D' bands upon IBS by Ar$^+$. Their intensities monotonically increase with the increase of $E_{ion}$ ($\geqslant$ 10 keV) for the same $\psi$, indicating that the ion-beam creates defects in the deeper region and/or more effectively for the larger $E_{ion}$. For the present sputtering condition, $E_{ion}$ is 2 keV,  small to critically damage the surface region, and leaves the sample in largely crystalline state.


\begin{figure*}[!t]
\begin{center}
\includegraphics [angle=0,width=0.45\textwidth]{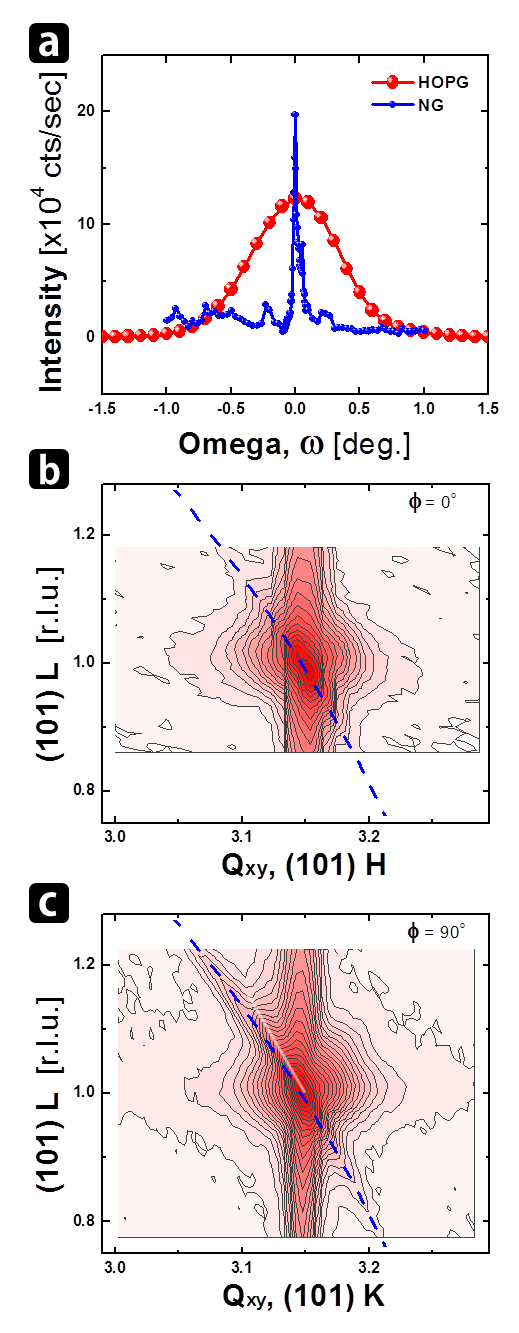}
\caption{(a) Spot profile around (006) peak of HOPG and natural graphite(NG). Reciprocal space maps (RSMs) around (101) peak of a $sputtered$ HOPG with $\psi$ = 5625 ions nm$^{-2}$, for Q$_{xy}$ (b) parallel to the ridge of the ripple and (c) perpendicular to the ridge. 
               }
\end{center}
\end{figure*}

Fig. 2(a) shows omega($\omega$)-rocking profiles of (0,0,6) peak  for both $sputtered$ HOPG and NG. The profile from the HOPG reveals a broad peak, indicating wide angular distribution of the grains. In contrast, NG shows sharp peaks that originate from large crystalline grains, indicating single crystal-like character of NG.  (See also AFM images in Figs. 3(a) and (b).)

\begin{table*}[h]
\renewcommand{\multirowsetup}{\centering}
\begin{tabular}{c|c|c|c|p{22mm}|p{22mm}}
\Xhline{3\arrayrulewidth}
\multirow{2}{15mm} & \multirow{2}{18mm}{Lattice a} & \multirow{2}{18mm}{Lattice c} & \multirow{2}{28mm}{Omega-rocking width} & \multicolumn{2}{c}{ Correlation length (00L)} \tabularnewline \cline{5-6}
 & & & & \centering Lateral & \centering Vertical \tabularnewline
 \hline \hline
HOPG & 2.4600 \AA & 6.7119 \AA & $ 0.75^{\circ} $ & 
\centering $ \sim 350 nm  $ &
\centering $ > 620 nm $  \tabularnewline
\hline
NG & 2.4605 \AA & 6.7082 \AA & $ \sim 0.01^{\circ} $  & \centering N/A & 
\centering $> 910 nm$ \tabularnewline
\Xhline{3\arrayrulewidth}
\end{tabular}
\caption{Lattice parameters and grain structures of $sputtered$ HOPG(0001) and NG(0001).}
\label{table1}
\end{table*}

Table 1 summarizes the lateral and vertical correlation lengths of both the $sputtered$ HOPG and NG. They are obtained by Williamson-Hall (WH) plots\cite{WHPlot}(plots not shown) of both peak width and peak rocking width versus the position of (0,0,L) peaks. 
The lateral correlation length of HOPG is $\sim$ 350 nm, comparable with the grain size observed in Fig. 3(a), a typical AFM image of a $pristine$ HOPG. The mean grain size of HOPG is still quite large compared with those of metallic films by an order of magnitude.\cite{D.Ghose2009, deMongeot} The lateral correlation length of NG cannot be determined by WH plots, because the peak width becomes smaller than the resolution limit of the detector as L of (0,0,L) peaks becomes small. This result indicates that the lateral correlation length of NG should be much larger than that of HOPG, as also suggested by the AFM images of both HOPG and NG respectively in Figs. 3(a) and (b). Each grain of NG should behave as single crystalline graphite.

Figs. 2 (b) and (c) show reciprocal space maps (RSMs) around (101) peak of a $sputtered$ HOPG with $\psi$ = 5625 ions nm$^{-2}$, for Q$_{xy}$ (reciprocal vector) (b) parallel to the ridge of the ripple and (c) perpendicular to the ridge. 
The blue dashed lines in both maps are equi-Q plots, originating from the angular (rotational) spread of the grain orientations analogous to the powder diffraction pattern. Note that a feature indicated by the white line is observed in the map perpendicular to the ripple, Fig. 2(c), but not for the map along the ripple. This is attributed to the formation of high index facets along the side wall of the ripples with the facet normal, $14.8 \pm 2.2^{\circ}$ from the global surface normal. 
Such a facet formation points to the crystalline surface layers, consistently with the conclusion from the Raman spectroscopy in Fig.1.

\begin{figure*}[!t]
\begin{center}
\includegraphics [angle=0,width=0.8\textwidth]{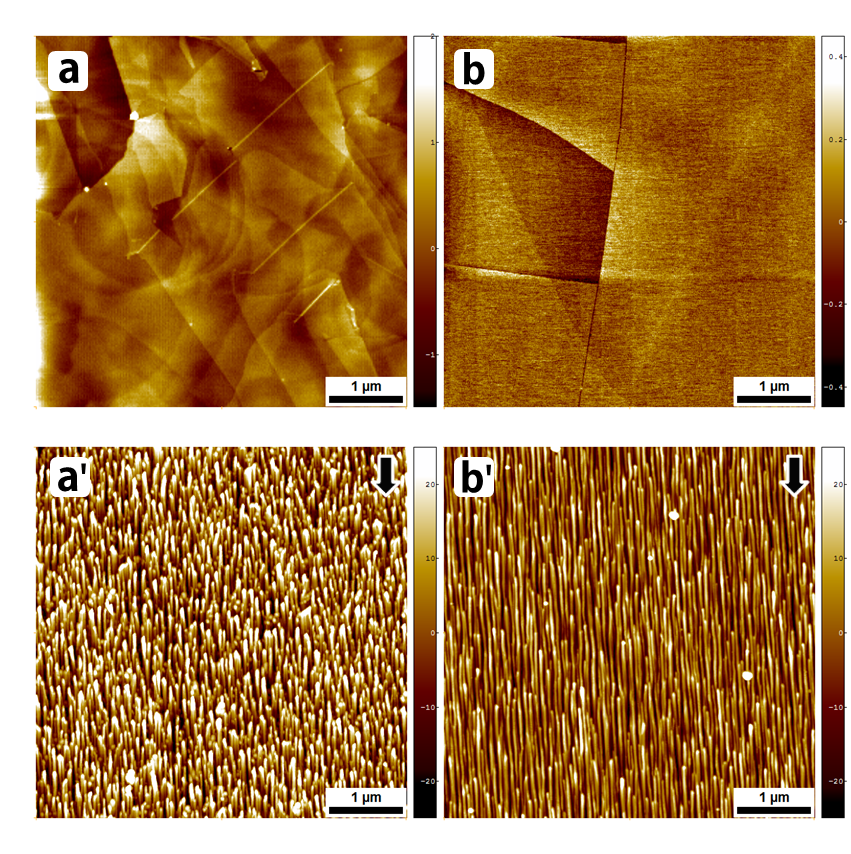}
\caption{ AFM images of both $pristine$ (a) HOPG and (b) NG (b). Nano ripple patterns on (a') HOPG(0001) and (b') NG(0001) after sputtering with $\psi = 5625$ ions nm$^{-2}$. (Image size: 5$\times 5$ $\mu m^2$) Arrow in each figure indicates the ion-beam direction. The unit of the z scale is nano meter. }
\end{center}
\end{figure*}



Figures 3(a') and (b') show ripple patterns on (a') HOPG(0001) and (b') NG(0001),  formed after extended IBS ($\psi = 5625$ ions nm$^{-2}$) under the same sputter condition. 
The mean uninterrupted ripple lengths or the coherence lengths $\ell$s show notable difference between HOPG and NG, 0.41 $\pm$ 0.06 $\mu$m for HOPG in Fig. 3(a'), much shorter than that for NG,  $ > 1.45 $ $\mu$m (not properly determined by the limited image size) in Fig. 3(b'). The error limit is set from the standard deviation of the measured $\ell$s that are obtained by manually measuring the uninterrupted lengths of the ripples along their ridges from the images. 
In addition, the surface width $W$ of the patterned HOPG in Fig. 3(a') is $\sim $13.46 nm,  notably larger than that of NG in Fig. 3(b'), $\sim $ 9.6 nm for the same sputter condition. That difference in $W$ is much larger than the difference between the initial surface widths of HOPG ($\sim 0.74$ nm) and NG ($\sim 0.43$ nm), and should have resulted from IBS. 
Since the conspicuous difference between the HOPG and NG lies in their mean grain sizes, the grained structure of the graphite 
should have affected both $\ell$ and $W$. 

In contrast, $\lambda$s are almost identical, $50 \pm 4.4$ nm and $50 \pm 3.4$ nm, respectively for HOPG and NG. (Taking line profiles from each image, we obtain the mean ripple wavelength $\lambda$.) The error limit is set from the standard deviation of the ripple wavelengths taken from the line profiles. $\lambda$s are smaller than the grain size, $ e. g. \sim$ 350 nm for HOPG.
Still 2 out of the 7 ripples of HOPG meet the grain boundary for HOPG, while the portion of the ripples neighboring the grain boundaries should be much smaller for NG due to much larger grain size than that for HOPG.  
The present observation that $\lambda$ is independent of the grain size indicates that the grain boundaries little affect coarsening of the ripple or that the grain boundaries parallel to the ridge direction do not hinder the mass transport across them. 
Similar conclusion is also drawn for polycrystalline metal films.\cite{TomaJAP,D.Ghose2009} 

\begin{figure*}[!t]
\begin{center}
\includegraphics [angle=0,width=0.8\textwidth]{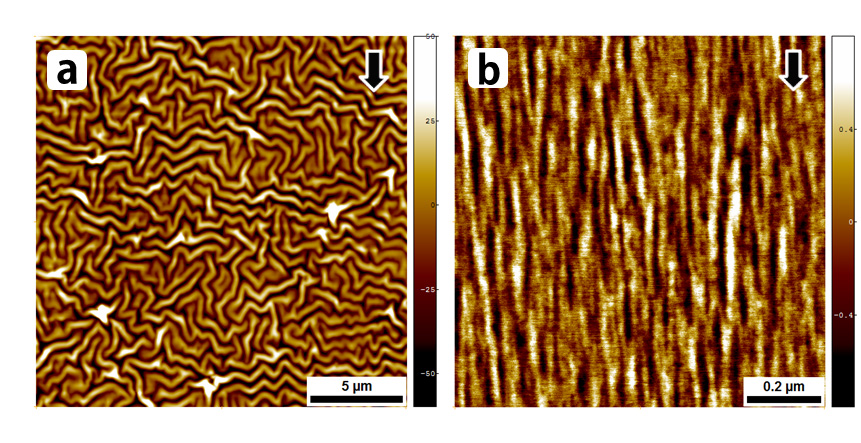}
\caption{ AFM images of nano patterns on (a) PDMS (Image size: 20$\times 20$ $\mu m^2$) and (b) Mica (size 1$\times 1$ $\mu m^2$) after sputtering under the same sputter conditions for  the graphites, with $\psi = 5625$ ions nm$^{-2}$. Arrow in each figure indicates the ion-beam direction. The unit of the z scale is nano meter. }
\end{center}
\end{figure*} 

A common feature of the ripples on both HOPG and NG is that the pronounced protuberances or the buds terminate ripples on the side facing the ion-beam. (Also see the inset of Fig. 1 and the following figures.) The other carbon allotropes such as diamond and tetrahedral amorphous carbon do not show such a bud structures upon IBS.\cite{Takahiro2007,Takahiro2009} Polymer(PDMS) show a labyrinthine pattern under the same sputter condition. (Fig. 4(a))  Layered structure may be a requirement for the formation of such protuberances. Muscovite mica, however, shows ripple pattern with no such protuberances under the same sputtering condition. (Fig. 4(b))  Mica is another layered material, but the interlayer bonding is stronger and constituent atoms are heaver than for graphite. Mica should then be less modified than graphite by the incidence of the same ion-beam. Since the pronounced protuberance needs high sputter yield producing large mass transport, we tentatively conclude that the bud termination of the ripples requires layered structure with weak interlayer bonding and/or light constituents of each layer. 

\section{Discussion}

\begin{figure*}[!t]
\begin{center}
\includegraphics [angle=0,width=0.8\textwidth]{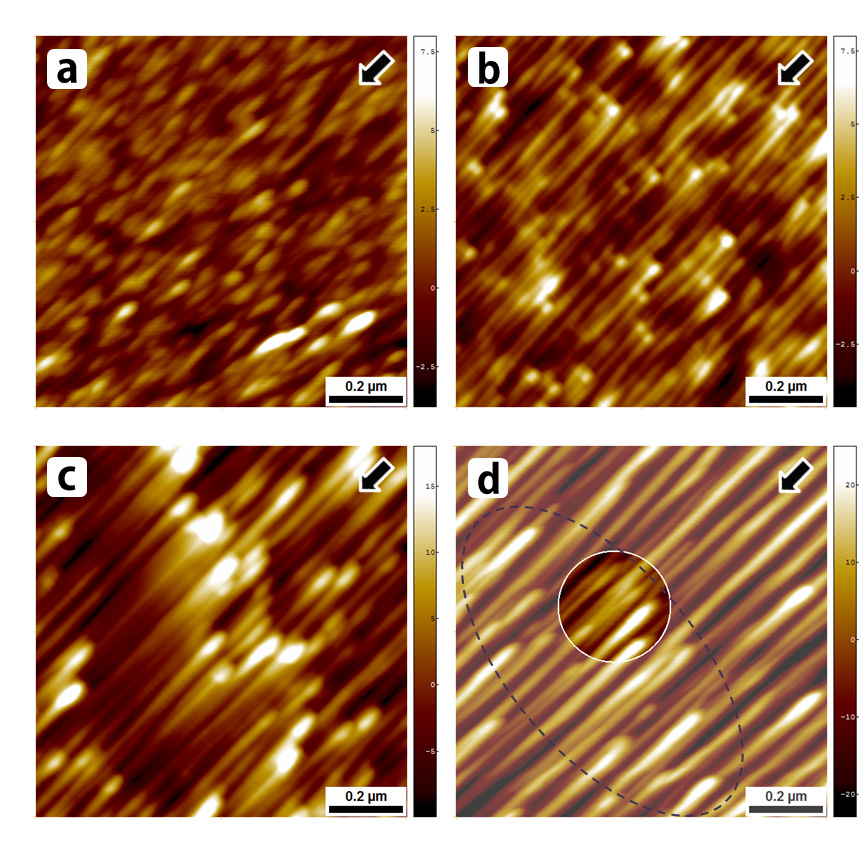}
\caption{ AFM images of nanopatterns on HOPG with the increase of $\psi$. (a) $\psi = 937$ ions nm$^{-2}$, (b) $\psi = 1875$ ions nm$^{-2}$, (c) $\psi = 2812$ ions nm$^{-2}$, (d) $\psi = 4687$ ions nm$^{-2}$. (Image size: 1$\times 1$ $\mu m^2$) Arrow in each figure indicates the ion-beam direction. The unit of the z scale is nano meter. }
\end{center}
\end{figure*}

The growth kinetics during the initial pattern formation on HOPG offers insights on how the grain boundaries influence the growth of the ripples in such a highly anisotropic fashion as observed in Fig. 3.
Fig. 5(a) shows an image at an incipient stage of pattern formation on HOPG with $\psi = 937$ ions nm$^{-2}$. 
The surface is covered by the oval-shaped islands elongated along the ion-beam direction. Some of the islands have already developed tails or the incipient ripples along the ion-beam direction. 
 The islands look distributed randomly, indicating little preference in the island formation between the grain boundaries and the terraces, which is also the case for NG. (Figure not shown) 

As sputtering proceeds with $\psi = 1875$ ions nm$^{-2}$, we observe that the islands have followed two different paths in their temporal evolution. (Fig. 5(b))
One kind of islands have grown to pronounced protuberances in the shape of buds as observed in Fig. 5(b). 
Note that the buds always form in the side of the ripple facing the ion-beam.
Their tails have also grown to form segmented ripples that are terminated by the adjacent buds along their ridges.
Each bud, thus, works as the birth place of the segmented ripple and simultaneously 
as the terminus of the adjacent ripple growing along the ion-beam direction. 
Ripple growth via elongation of the tails has also been observed for the polycrystalline Ni films.\cite{Skeren}

The other kind of islands become affiliated with the neighboring ripples, and barely idetifiable as nodes
along their ridges.(Fig. 5(b))  
Only the islands in the form of the buds remain distinguished, and the density of the islands looks significantly diminished as compared with that in Fig. 5(a).
Note that the growth of the ripple occurs via the linkage-which can occur multiply-of the ripples along the ion-beam direction, for which
 the nodes work as linker. 
Due to the linkage of the neighboring ripples that do not meet along a line, the elongated ripples slightly meander throughout their length as shown in Fig. 5(b) and also in Fig. 3(b').

With the continued sputtering with $\psi = 2812$ ions nm$^{-2}$, each ripple in Fig. 5(c) has grown longer, wider and shows the less modulation in both the height and width than the ripples in Fig. 5(b). The buds have also become further enlarged. Contrastingly, the nodes  are not identified any more along the ridge, indicating the the mass redistribution around each node is very efficient.
With the further sputtering with $\psi = 4687$ ions nm$^{-2}$, ripples are so coarse that adjacent ripples are almost in touch. As a result, the side-by-side coalescence of adjacent ripples is frequently observed as seen in the circled area in Fig. 5(d).
Such coalescence of the ripples  improves the order of the ripple pattern, because 
IBS drives the coalescence of the adjacent ripples, leading to the alignment of the merged ripples along the ion-beam direction. 
Stepanova {\it et al.}\cite{Stepanova} also observe in their kinetic Monte Carlo simulation
that the coalescence improves the order of the ripples  by eliminating defects in the pattern.   

Note that the distribution of the buds is not so random in Fig. 5(d) as that of the incipient islands in Fig. 5(a). 
The buds often locally group or separated by region depleted of the buds, conspicuously in Fig. 5(c) and also in Fig. 5(d). 
Such grouping of the islands are more and more frequently observed with the larger $\psi$. The distance between the two neighboring groups in Figs. 5(c) and (d) is around 400 nm that is similar to the coherence length $\ell$ of the ripple pattern and also the mean grain size.  This suggests the close correlation between the grouping of the islands and the grained structure.  

From the temporal evolution of the patterns in Fig. 5, the growth kinetics of the two kinds of primordial islands can be summarized as follows; The $first$ kind of islands or 1) the buds form only on the side of each ripple facing the ion-beam. 2) The bud grows up as sputter proceeds. 3) The buds tend to locally group. 4) The mean distance between the neighboring buds along the ion-beam direction is similar to the mean grain size. 
 The $second$ kind of islands or 1) the nodes reside in the middle of the ripples or each of the nodes stay inbetween two neighboring buds defining a ripple. 2) The nodes swiftly die out as sputter proceeds, indicating efficient diffusion processes of adspecies around them.  

Upon those observation, we propose the following picture for the ripple growth on HOPG(0001); The buds form at the grain boundaries, while the nodes on the terraces of the grains. 
In the early stage of sputtering, the grain boundaries are not influential on the formation of the primordial islands because the roughness of the $pristine$ surface is of the atomic scale as shown in Figs. 3(a) and (b). With the continued sputtering, however, the initially embedded side walls of the grains become exposed. 
The side wall-formed of the edges of the graphenes-is more easily modified by the incident ion-beam than the surface of the graphite surface. In addition, the ion-beam shed the enhanced ion flux on the side walls, especially those facing the incident ion-beam due to the reduced angle from the local surface normal, and there. 
With all the effects in synergy,  
the growth of the islands  becomes escalated near the grain boundaries facing the incident ion-beam, 
and the buds selectively decorate them.  

On the other hand, the diffusion of adspecies  is highly efficient on the terrace due to strong intraplanar $sp^2$ bond of the graphite surface.\cite{carbon diffusion,Flake diffusion}  Then, the islands on the terrace becomes linked to the adjacent ripple to form the nodes as seen in Fig. 5(b) that becomes swiftly transformed to be a part of  the respective, now elongated ripple as in Figs. 5(c) and 5(d).
In the long run, each ripple extends via the linkage of the ripples from a bud on a grain boundary facing the ion-beam to the other bud on the adjacent grain boundary. $\ell$ of the ripple pattern should, thus, be congruent to the mean grain size, as actually observed in the experiments.  

The present picture can also elucidate the experimental observation that the surface width $W$ of the patterned HOPG is always larger than that of NG for the same sputter condition. In Fig. 3, for example, $ W$ of the patterned HOPG is $\sim13.46$ nm, larger than that of NG,  $\sim 9.6$ nm. The mean height of the nano buds from the adjacent ridge of the ripple is 9.9 $\pm$ 0.7 nm for both HOPG and NG, and is a major source of the surface width. Since the density of the nano buds is higher for HOPG due to the smaller grain size than that for NG, $W$ is, thus, larger for HOPG than for NG. This observation assures the significant role of the grain boundaries in the morphological evolution of the ripple pattern.

Experimental observation of the similar $\lambda$s for both HOPG and NG in Fig. 2  indicates that the grain boundaries 
do not affect the coarsening of the ripple. 
Under the present experimental condition, the mass transport across the grain boundary, transversally to the ion-beam is not driven by thermal diffusion that should be seriously hampered  by the grain boundary. Instead, athermal processes such as sputter-induced 1) solid flow\cite{Umbach,Castro2012} and/or 2) ballistic diffusion\cite{Carter,Norris2011} must play the major roles. Such athermal diffusion could be substantial for graphite, because 1) the small mass of constituent of HOPG or carbon compared with that of the projectile or argon leads to large momentum transfer from the incident Ar ion. Habenicht\cite{Habenicht2001} observed linear dependence of $\lambda$ on incident ion energy, consistent with a prediction for the ion-induced solid flow for HOPG\cite{Castro2012} Recently, amorphous carbon sputtered by Xe$^{+}$ of 0.2 to 10 keV showed significant mass redistribution by the incident ion, which was also found the major source of the surface instability.\cite{Hofsass2012} Those results support the argument of the significant contribution from the sputter-induced mass transport to the pattern formation on HOPG(0001).  

Recently, \v{S}kere\v{n} {\it et al.}\cite{Skeren} proposed a model for the pattern formation of polycrystalline (Ni) films. The variation of the sputter yield  depending on the orientations of the grains forming the film can well reproduce the observed pattern evolution on Ni films formed under various sputter conditions without invoking the well-known instability due to curvature dependent erosion of surface by IBS.\cite{BH} The $pristine$ HOPG is largely formed of the grains differing only in relative azimuthal angle, but sharing (0001) basal plane with minute mosaic 
estimated from the omega-rocking width in Table I. The sputter yields among grains should not  significantly differ. With continued sputtering, however, the side walls at the grain boundaries in the direction facing the ion-beam becomes exposed, and have distinctly larger sputter yield than (0001) basal plane. This-dynamically driven-inhomogeneous surface with respect to the sputter yield  is similar to the situation for the polycrystalline Ni film, and actually leads to an instability or to the formation of the buds near the grain boundaries. This is an aspect in common with the model of \v{S}kere\v{n} {\it et al.}\cite{Skeren}. Away from the grain boundary or on terrace, however, the ripple formation should follow the same mechanism for both HOPG and NG as expected for single crystalline surface.  

In the previous work with the polycrystalline Au films\cite{deMongeot}, the large height fluctuation of Au grains leads to shadowing instability. 
The present HOPG(0001) is distinct from the metallic film, because the lateral grain size is much larger than the characteristic wavelength and the initial surface width is very small, less than 1 nm. Due to the small surface height fluctuation, the shadowed area is negligible. For the Ni film whose initial surface width is also less than 1 nm, \v{S}kere\v{n} {\it et al.}\cite{Skeren} also observe the rapid reduction of the shaded area as sputter proceeds, and the negligible effects of shadowing.  Accordingly, the temporal evolution of HOPG do not show any sign of the initial smoothening observed for the Au film, but show monotonic increases of both $W$.\cite{unpublished}  
Instead, the grain boundaries play significant role in the determination of both $\ell$ and $W$ of the ripple pattern  in contrast to the case of the Au films.



\section{Summary and conclusions}
 
  We investigated the effects of the grained structure on the pattern formation by ion-beam-sputtering (IBS), employing two different graphites: highly oriented pyrolytic graphite (HOPG) and natural graphite (NG) whose mean grain size is distinctly larger than that of HOPG. 
Each ripple runs from a pronounced protuberance or a bud at one grain boundary to that at the adjacent boundary.  
 The buds originate from the side walls of the grains-formed of the edges of the graphenes, thus having high sputter yield- that become exposed at the grain boundaries facing the ion-beam by continued sputtering. 
  
  Since HOPG is composed of the smaller grains than NG, so is the mean uninterrupted ripple length of HOPG.
  Due to the higher bud density on HOPG, the surface width of the patterned HOPG is larger than that of NG, 
  well elucidating the experimental observations. 
  On the other hand, the wavelengths on both HOPG and NG are similar, indicating that the ripple coarsening across the grain boundary proceeds via athermal processes such as sputter-induced viscous flow or ballistic diffusion.\cite{Umbach,Castro2012,Hofsass2012,Norris2011}  
In short, the grain boundary of the graphite significantly affects the morphological evolution of the ripple pattern on graphite, 
but in highly anisotropic fashion.

{\it Acknowledgments} This work was supported in part by NRF grant (JK, 2013030406). JK appreciates R. Cuerno for comments and discussion.

\end{document}